%% file: gang16.tex
\documentclass[conference]{IEEEtran}
\usepackage[cmex10]{amsmath}
\usepackage{url}
\usepackage{array}
\usepackage{color}
\usepackage{graphicx}
\usepackage{mdwmath}
\usepackage{mdwtab}
\usepackage{eqparbox}
\graphicspath{ {images/} }
\usepackage{subcaption}
\usepackage{multirow}
\usepackage{textcomp}
\usepackage{comment}
\usepackage{lipsum}

\IEEEoverridecommandlockouts

\usepackage{fancyhdr} 
\usepackage{kantlipsum} 
\fancypagestyle{plain}{ 
\fancyhf{} 
\fancyhead[C]{Conference on \LaTeX} 
 
}

\usepackage{eso-pic} 
\begin{document} 

\newenvironment{itemNoSpace}{
\begin{itemize}
  \setlength{\itemsep}{1pt}
  \setlength{\parskip}{0pt}
  \setlength{\parsep}{0pt}}{\end{itemize}
}

\pdfinfo{
   /Author (Lakshika Balasuriya, Sanjaya Wijeratne, Derek Doran, Amit Sheth)
   /Title (Finding Street Gang Members on Twitter)
   /Keywords (Street Gangs;Twitter Profile Identification;Gang Activity Understanding;Social Media Analysis)
   /Subject (Finding Street Gang Members on Twitter - ASONAM 2016 Conference Paper)
}

\title{Finding Street Gang Members on Twitter}
\author{\IEEEauthorblockN{Lakshika Balasuriya}
\IEEEauthorblockA{Kno.e.sis Center\\Wright State University\\
Dayton OH, USA\\
lakshika@knoesis.org}
\and
\IEEEauthorblockN{Sanjaya Wijeratne}
\IEEEauthorblockA{Kno.e.sis Center\\Wright State University\\
Dayton OH, USA\\
sanjaya@knoesis.org}
\and
\IEEEauthorblockN{Derek Doran}
\IEEEauthorblockA{Kno.e.sis Center\\Wright State University\\
Dayton OH, USA\\
derek@knoesis.org}
\and
\IEEEauthorblockN{Amit Sheth}
\IEEEauthorblockA{Kno.e.sis Center\\Wright State University\\
Dayton OH, USA\\
amit@knoesis.org}
}
\maketitle

\begin{abstract}

Most street gang members use Twitter to intimidate others, to present outrageous images and statements to the world, and to share recent illegal activities. Their tweets may thus be useful to law enforcement agencies to discover clues about recent crimes or to anticipate ones that may occur. Finding these posts, however, requires a method to discover gang member Twitter profiles. This is a challenging task since gang members represent a very small population of the 320 million Twitter users. This paper studies the problem of automatically finding gang members on Twitter. It outlines a process to curate one of the largest sets of verifiable gang member profiles that have ever been studied. A review of these profiles establishes differences in the language,  images,  YouTube links, and  emojis gang members use compared to the rest of the Twitter population. Features from this review are used to train a series of supervised classifiers. Our classifier achieves a promising $F_1$ score with a low false positive rate. 
\end{abstract}


\section{Introduction and Motivation}

The crime and violence street gangs introduce into neighborhoods is a growing epidemic in cities around the world\footnote{\url{http://goo.gl/OjWeYf}}. Today, over 1.23 million people in the United States are members of a {\em street gang}~\cite{2011NationalGangThreat, 2013NationalGangReport}, which is a coalition of peers, united by mutual interests, with identifiable leadership and internal organization, who act collectively to conduct illegal activity and to control a territory, facility, or enterprise~\cite{miller1992crime}. They promote criminal activities such as drug trafficking, assault, robbery, and threatening or intimidating a neighborhood~\cite{2013NationalGangReport}. Moreover, data from the Centers for Disease Control in the United States suggests that the victims of at least 1.3\% of all gang-related\footnote{The terms `gang' and `street gang' are used interchangeably in this paper.} homicides are merely innocent bystanders who live in gang occupied neighborhoods~\cite{centers2012gang}.

Street gang members have established online presences coinciding with their physical occupation of neighborhoods. The National Gang Threat Assessment Report confirms that at least tens of thousands of gang members are using social networking websites such as Twitter and video sharing websites such as YouTube in their daily life~\cite{2011NationalGangThreat}. They are very active online; the 2007 National Assessment Center's survey of gang members found that 25\% of individuals in gangs use the Internet for at least 4 hours a week~\cite{2007assesment}. Gang members typically use social networking sites and social media to develop online respect for their street gang~\cite{King2007S66} and to post intimidating, threatening images or videos~\cite{doi:10.1080/07418825.2013.778326}. This ``Cyber-'' or ``Internet banging''~\cite{patton13} behavior is precipitated by the fact that an increasing number of young members of the society are joining gangs~\cite{howell2010gang}, and these young members have become enamored with technology and with the notion of sharing information quickly and publicly through social media\footnote{\url{http://www.hhs.gov/ash/oah/news/e-updates/eupdate-nov-2013.html}}. Stronger police surveillance in the physical spaces where gangs congregate further encourages gang members to seek out virtual spaces such as social media to express their affiliation, to sell drugs, and to celebrate their illegal activities~\cite{ito10}.

Gang members are able to post publicly on Twitter without fear of consequences because there are few tools law enforcement can use to surveil this medium~\cite{7165945}. Police departments across the United States instead rely on manual processes to search social media for gang member profiles and to study their posts. For example, the New York City police department employs over 300 detectives to combat teen violence triggered by insults, dares, and threats exchanged on social media, and the Toronto police department teaches officers about the use of social media in investigations~\cite{police13}. Officer training is broadly limited to understanding policies on using Twitter in investigations and best practices for data storage~\cite{brunty14}. The safety and security of city neighborhoods can thus be improved if law enforcement were equipped with intelligent tools to study social media for gang activity.

\IEEEpubidadjcol 
The need for better tools for law enforcement cannot be underscored enough. Recent news reports have shown that many incidents involving gangs start on Twitter, escalate over time, and lead to an offline event that could have been prevented by an early warning. For example, the media reported on a possible connection between the death of a teenage rapper from Illinois and the final set of tweets he posted. One of his last tweets linked to a video of him shouting vulgar words at a rival gang member who, in return, replied {\em``I'ma kill you''} on social media\footnote{\url{http://www.wired.com/2013/09/gangs-of-social-media/}}. In a following tweet, the teenage rapper posted {\em ``im on 069''}, revealing his location, and was shot dead soon after that post. Subsequent investigation revealed that the rivalry leading to his death began and was carried out entirely on social media. Other reporting has revealed how innocent bystanders have also become targets in online fights, leaving everyone in a neighborhood at risk\footnote{\url{https://goo.gl/75U3ME}}.

\begin{figure}
\centering
\includegraphics[width=1.0\linewidth]{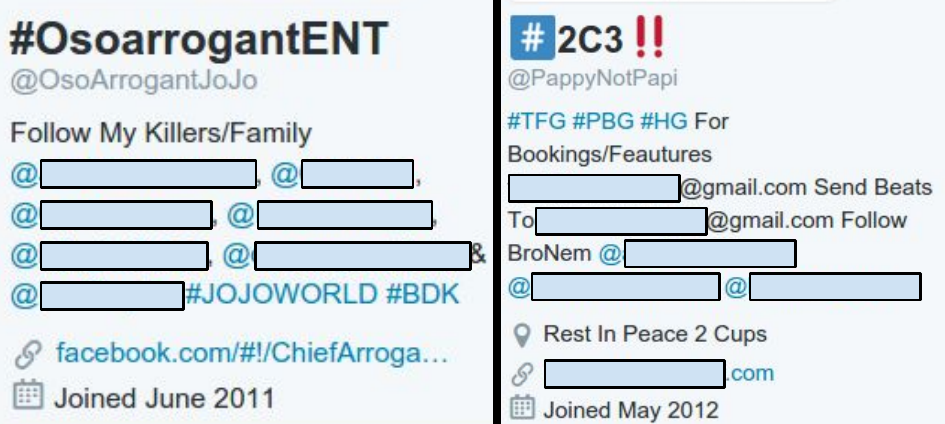} 
\caption{Twitter profile descriptions of known gang members. Pursuant to an IRB governing human subject research, we are prohibited from revealing personally identifiable information in this paper. We only report Twitter handles that have already been revealed in widely reported publications and were not collected by the research team for this work.}
\label{fig:twitterprofiles}
\vspace{-20px}
\end{figure}

This paper investigates whether gang member profiles can be identified automatically on Twitter, which can enable better surveillance of gang members on social media. Classifying Twitter profiles into particular types of users has been done in other contexts~\cite{pennacchiotti2011machine, tinati2012identifying, liu2013s}, but gang member profiles pose unique challenges. For example, many Twitter profile classifiers search for contextual clues in tweets and profile descriptions~\cite{purohit2012user}, but gang member profiles use a rapidly changing lexicon of keywords and phrases that often have only a local, geographic context. This is illustrated in Figure~\ref{fig:twitterprofiles}, which shows the Twitter profile descriptions of two verified deceased gang members. The profile of @OsoArrogantJoJo provides evidence that he belongs to a rival gang of the Black Disciples by \texttt{\#BDK}, a hashtag that is only known to those involved with gang culture in Chicago. @PappyNotPapi's profile mentions \texttt{\#PBG} and our investigations revealed that this hashtag is newly founded and stands for the Pooh Bear Gang, a gang that was formerly known as the Insane Cutthroat Gangsters. Given the very local, rapidly changing lexicon of gang members on social media, building a database of keywords, phrases, and other identifiers to find gang members nationally is not feasible. Instead, this study proposes heterogeneous sets of features derived not only from profile and tweet text but also from the emoji usage, profile images, and links to YouTube videos reflecting their music culture. A large set of gang member profiles, obtained through a careful data collection process, is compared against non-gang member profiles to find contrasting features. Experimental results show that using these sets of features, we can build a classifier that has a low false positive rate and a promising $F1$-score of 0.7755.

This paper is organized as follows. Section~\ref{sec:rr} discusses the related literature and positions how this work differs from other related works. Section \ref{sec:dc} discusses the data collection, manual feature selection and our approach to identify gang member profiles. Section \ref{sec:eval} gives a detailed explanation for evaluation of the proposed method and the results in detail. Section \ref{sec:con} concludes the work reported while discussing the future work planned.

\section{Related Work} \label{sec:rr}

Gang violence is a well studied social science topic dating back to 1927~\cite{thrasher1963gang}. However, the notions of ``Cyber-'' or ``Internet banging'', which is defined as {\em {``the phenomenon of gang affiliates using social media sites to trade insults or make violent threats that lead to homicide or victimization''}}~\cite{patton13}, was only recently introduced~\cite{patton2014social,7165945}. Patton {\em et al.} introduced the concept of ``Internet banging'' and studied how social media is now being used as a tool for gang self-promotion and as a way for gang members to gain and maintain street credibility~\cite{patton13}. They also discussed the relationship between gang-related crime and hip-hop culture, giving examples on how hip-hop music shared on social media websites targeted at harassing rival gang members often ended up in real-world collisions among those gangs. Decker {\em et al.} and Patton {\em et al.} have also reported that street gangs perform Internet banging with social media posts of videos depicting their illegal behaviors, threats to rival gangs, and firearms~\cite{decker2011leaving,desmondupatton2015}.

The ability to take action on these discoveries is limited by the tools available to discover gang members on social media and to analyze the content they post~\cite{patton2014social}. Recent attempts to improve our abilities include a proposed architecture for a surveillance system that can learn the structure, function, and operation of gangs through what they post on social media~\cite{7165945}. However, the architecture requires a set of gang member profiles for input, thus assuming that they have already been discovered. Patton {\em et al.}~\cite{desmondupatton2015} devised a method to automatically collect tweets from a group of gang members operating in Detroit, MI. However, their approach required the profile names of the gang members to be known beforehand, and data collection was localized to a single city in the country.

This work builds upon existing methods to automatically discover gang member
profiles on Twitter. This type of user profile classification problem has been explored in a diverse set of applications such as political affiliation~\cite{pennacchiotti2011machine}, ethnicity~\cite{pennacchiotti2011machine}, gender~\cite{liu2013s}, predicting brand loyalty~\cite{pennacchiotti2011machine}, and user occupations~\cite{purohit2012user}. However, these approaches may utilize an abundance of positive examples in their training data, and only rely on a single feature type (typically, tweet text). Whereas most profile classifiers focus on a single {\em type} of feature (e.g. profile text), we consider the use of a variety of feature types, including emoji, YouTube links, and photo features.

\begin{figure}
\centering
\includegraphics[width=1\linewidth]{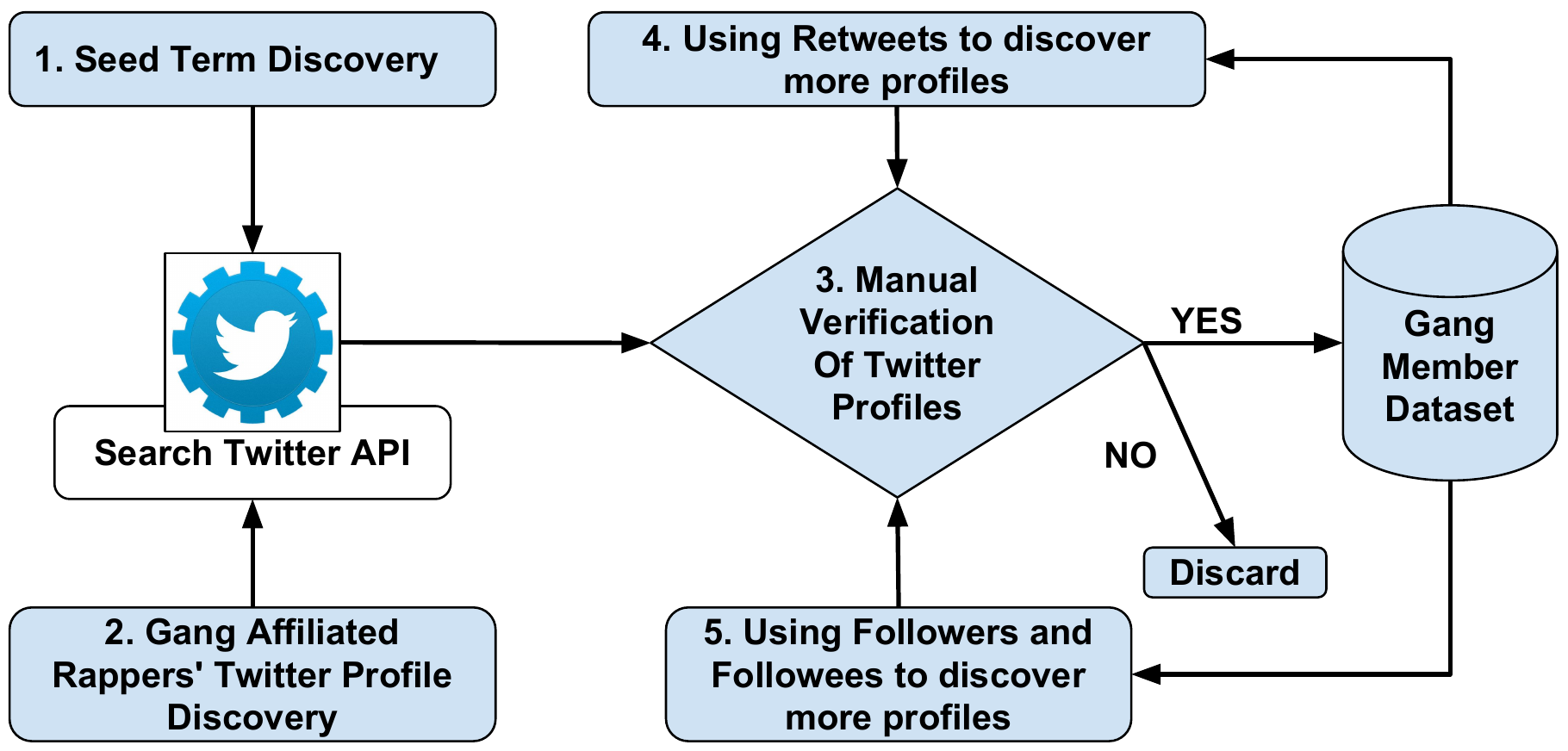} 
\caption{Gang member dataset creation.}
\label{fig:workflow}
\vspace{-20px}
\end{figure}

\section{Discovering Gang Member Profiles} \label{sec:dc}

This section discusses the methodology we followed to study and classify the Twitter profiles of gang members automatically. It includes a semi-automatic data collection process to discover a large set of verifiable gang member profiles, an evaluation of the tweets of gang and non-gang member posts to identify promising features, and the deployment of multiple supervised learning algorithms to perform the classification.

\begin{table}[h]
\begin{center}
\begin{tabular}{ |l | c| }
\hline
\textbf{Method}& \textbf{Number of Profiles}\\\hline
 Seed term discovery & 280 \\\hline
 Gang Affiliated Rappers &22 \\\hline
 Retweets, Followers \& Followees &98 \\\hline
 \textbf{Total} &\textbf{400} \\\hline
\end{tabular}
\end{center}
\caption{Number of gang member profiles captured.}
\label{datacollection:approach}
\vspace{-20px}
\end{table}

\subsection{Data collection}

Discovering gang member profiles on Twitter to build training and testing datasets is a challenging task. Past strategies to find these profiles were to search for keywords, phrases, and events that are known to be related to gang activity in a particular city a priori~\cite{7165945,desmondupatton2015}. However, such approaches are unlikely to yield adequate data to train an automatic classifier since gang members from different geographic locations and cultures use local languages, location-specific hashtags, and share information related to activities in a local region~\cite{7165945}. Such region-specific tweets and profiles may be used to train a classifier to find gang members within a small region but not across the Twitterverse. To overcome these limitations, we adopted a semi-automatic workflow, illustrated in Figure~\ref{fig:workflow}, to build a dataset of gang member profiles suitable for training a classifier. The steps of the workflow are:

\noindent {\bf 1. Seed Term Discovery:}
Following the success of identifying gang member profiles from Chicago~\cite{7165945}, we began our data collection with discovering universal terms used by gang members. We first searched for profiles with hashtags for Chicago gangs noted in~\cite{7165945}, namely \texttt{\#BDK} (Black Disciple Killers) and \texttt{\#GDK} (Gangster Disciples Killers). Those profiles were analyzed and manually verified as explained in Step 3. Analysis of these profiles identified a small set of hashtags they all use in their profile descriptions. Searching Twitter profiles using those hashtags, we observed that gang members across the U.S. use them, thus we consider those terms to be location neutral. For example, gang members post \texttt{\#FreeDaGuys} in their profile to support their fellow members who are in jail, \texttt{\#RIPDaGuys} to convey the grieving for fallen gang members, and \texttt{\#FuckDaOpps} to show their hatred towards police officers. We used these terms as keywords to discover Twitter profiles irrespective of geographical location. We used the Followerwonk Web service API\footnote{\url{https://moz.com/followerwonk/bio}} and Twitter REST API\footnote{\url{https://dev.twitter.com/rest/public}} to search Twitter profile descriptions by keywords \texttt{\#FreeDaGuys}, \texttt{\#FreeMyNigga}, \texttt{\#RIPDaGuys}, and \texttt{\#FuckDaOpps}. Since there are different informal ways people spell a word in social media, we also considered variations on the spelling of each keyword; for example, for \texttt{\#FreeDaGuys}, we searched both \texttt{\#FreeDaGuys}, and \texttt{\#FreeTheGuys}.

\noindent {\bf 2. Gang Affiliated Rappers' Twitter Profile Discovery:} 
Finding profiles by a small set of keywords is unlikely to yield sufficient data. Thus, we sought additional gang member profiles with an observation from Patton {\em et al.}~\cite{patton13} that the influence of hip-hop music and culture on offline gang member activities can also be seen in their social media posts. We thus also consider the influence of hip-hop culture on Twitter by exploring the Twitter network of known gangster rappers who were murdered in 2015 due to gang-related incidents\footnote{\url{http://www.hipwiki.com/List+of+Rappers+Murdered+in+2015}}. We searched for these rapper profiles on Twitter and manually checked that the rapper was affiliated to a gang.

\noindent {\bf 3. Manual verification of Twitter profiles:} 
We verified each profile discovered manually by examining the profile picture, profile background image, recent tweets, and recent pictures posted by a user. During these checks, we searched for terms, activities, and symbols that we believed could be associated with a gang. For example, profiles whose image or background included guns in a threatening way, stacks of money, showing gang hand signs and gestures, and humans holding or posing with a gun, appeared likely to be from a gang member. Such images were often identified in profiles of users who submitted tweets that contain messages of support or sadness for prisoners or recently fallen gang members, or used a high volume of threatening and intimidating slang language. Only profiles where the images, words, and tweets all suggested gang affiliation were labeled as gang affiliates and added to our dataset. Although this manual verification does have a degree of subjectivity, in practice, the images and words used by gang members on social media are so pronounced that we believe any reasonable analyst would agree that they are gang members. We found that not all the profiles collected belonged to gang members; we observed relatives and followers of gang members posting the same hashtags as in Step 1 to convey similar feelings in their profile descriptions.

\noindent {\bf 4. Using Retweets to discover more profiles:} 
From the set of verified profiles, we explored their retweet and follower networks as a way to expand the dataset. We first considered authors of tweets which were retweeted by a gang member in our seed set. In Twitter, {\em``retweeting''} is a mechanism by which a user can share someone else's tweet to their follower audience. Assuming that a user only retweets things that they believe or their audience would be interested in, it may be reasonable to assume that gang members would only be interested in sharing what other gang members have to say, and hence, the authors of gang members' retweets could also be gang members.

\noindent {\bf 5. Using Followers and Followees to discover more profiles:}
We analyzed followers and followees of our seed gang member profiles to find more gang member profiles. A Twitter user can follow other Twitter users so that the individual will be subscribed to their tweets as a follower and they will be able to start a private conversation by sending direct messages to the individual. Motivated by the sociological concept of homophily, which claims that individuals have a tendency to associate and bond with similar others\footnote{\url{http://aris.ss.uci.edu/~lin/52.pdf}}, we hypothesized that the followers and followees of Twitter profiles from the seed set may also be gang members. Manual verification of Twitter profiles collected from retweets, followers, and followees of gang members showed that a majority of those profiles are non-gang members who are either family members, hip-hop artists, women or profiles with pornographic content. To ensure that our dataset is not biased towards a specific gang or geographic location, only a limited number of profiles were collected via retweets, followers and followees. 

\begin{figure}[ht]
\begin{subfigure}[b]{0.45\linewidth}
\includegraphics[width=1.15\linewidth, height=1.25\linewidth]{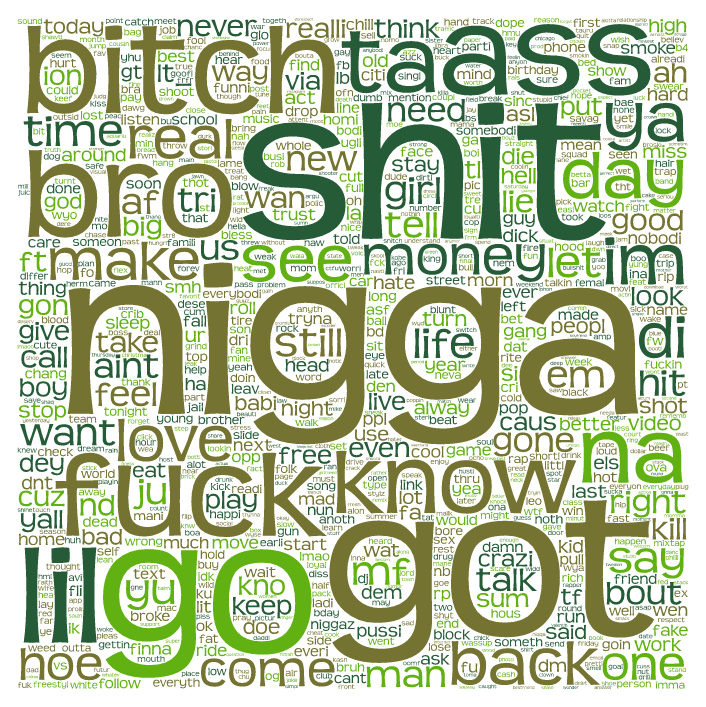} 
\caption{Gang members.}
\label{fig:subim111}
\end{subfigure}
\quad
\begin{subfigure}[b]{0.45\linewidth}
\includegraphics[width=1.15\linewidth, height=1.25\linewidth]{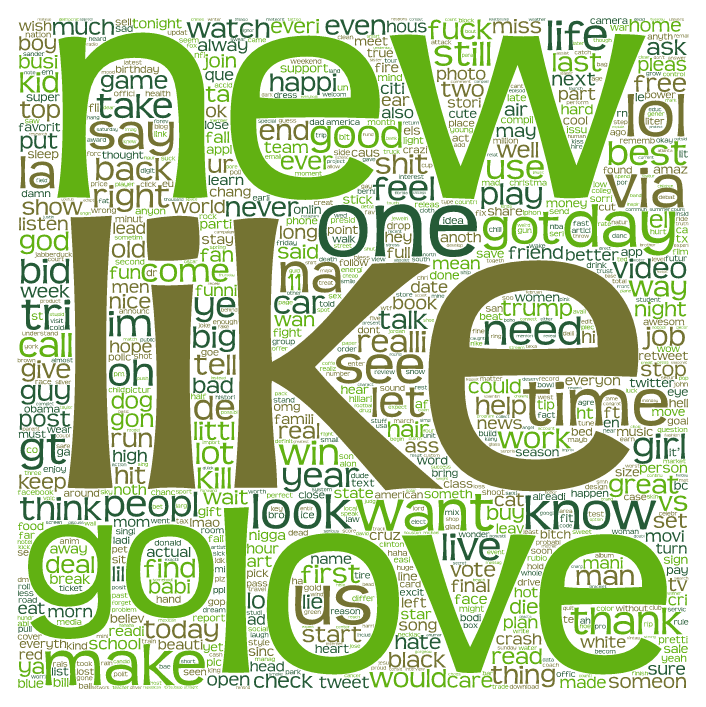}
\caption{Non-gang members.}
\label{fig:subim222}
\end{subfigure}
\caption{Comparison of words used in tweets.}
\label{fig:image2}
\vspace{-20px}
\end{figure}

\noindent 
Table~\ref{datacollection:approach} summarizes the number of profiles manually verified as gang members from Twitter profiles collected in step 1, 2, 4 and 5. Altogether we collected 400 gang member's Twitter profiles. This is a large number compared to previous studies of gang member activities on social media that curated a maximum of 91 profiles~\cite{7165945}. Moreover, we believe the profiles collected represent a diverse set of gang members that are not biased toward a particular geographic area or lingo as our data collection process used location-independent terms proven to be used by gang members when they express themselves.

\subsection{Data analysis} \label{sec:data_analysis} 
We next explore differences between gang and non-gang member Twitter usage to find promising features for classifying profiles. For this purpose, profiles of non-gang members were collected from the Twitter Streaming API\footnote{\url{https://dev.twitter.com/streaming/overview}}. We collected a random sample of tweets and the profiles of the users who authored the tweets in the random sample. We manually verified that all Twitter profiles collected in this approach belong to non-gang members. The profiles selected were then filtered by location to remove non-U.S. profiles by reverse geo-coding the location stated in their profile description by the Google Maps API\footnote{\url{https://developers.google.com/maps/}}. Profiles with location descriptions that were unspecified or did not relate to a location in the U.S. were discarded. We collected 2,000 non-gang member profiles in this manner. In addition, we added 865 manually verified non-gang member profiles collected using the location neutral keywords discussed in Section~\ref{sec:dc}. Introducing these profiles, which have some characteristics of gang members (such as cursing frequently or cursing at law enforcement) but are not, captures local languages used by family/friends of gang members and ordinary people in a neighborhood where gangs operate.

With the Twitter REST API\footnote{\url{https://dev.twitter.com/rest/public}}, we collected the maximum number of most recent tweets that can be retrieved (3,200) along with profile descriptions and images (profile and cover photos) of every gang and non-gang member profile. The resulting dataset consists of 400 gang member Twitter profiles and 2,865 non-gang member Twitter profiles. The dataset has a total of 821,412 tweets from gang member profiles and 7,238,758 tweets from non-gang member profiles. Prior to analyzing any text content, we removed all of the seed words used to find gang member profiles, all stop words, and performed stemming across all tweets and profile descriptions.

\begin{figure}
\centering
\includegraphics[width=0.95\linewidth]{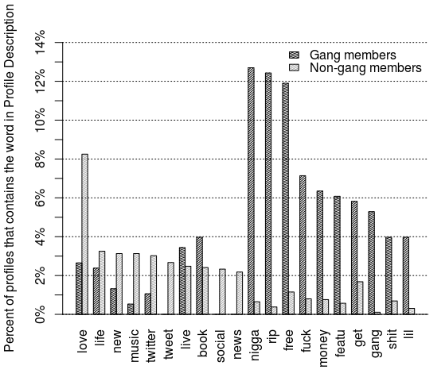} 
\caption{Word usage in profile descriptions: gang vs non-gang.}
\label{profile_description}
\vspace{-20px}
\end{figure}

\subsubsection{Tweet text} \label{sec:data_analysis_tweet_text}
Figure~\ref{fig:image2} summarizes the words seen most often in the gang and non-gang members' tweets as clouds. They show a clear difference in language. For example, we note that gang members more frequently use curse words in comparison to ordinary users. Although cursing is frequent in tweets, they represent just 1.15\% of all words used~\cite{wang2014cursing}. In contrast, we found 5.72\% of all words posted by gang member accounts to be classified as a curse word, which is nearly five times more than the average curse word usage on Twitter. The clouds also reflect the fact that gang members often talk about drugs and money with terms such as {\em smoke, high, hit}, and {\em money}, while ordinary users hardly speak about finances and drugs. We also noticed that gang members talk about material things with terms such as {\em got, money, make, real, need} whereas ordinary users tend to vocalize their feelings with terms such as {\em new, like, love, know, want, look, make, us}. These differences make it clear that the individual words used by gang and non-gang members will be relevant features for gang profile classification.

\subsubsection{Twitter Profile Description} 
On Twitter, a user can give a self-description as a part of the user's profile. A comparison of the top 10 words in gang members' and non-gang members' Twitter profile descriptions is shown in Figure~\ref{profile_description}. The first 10 words are the most frequently used words in non-gang members' profiles and the latter 10 words are the most frequently used words in gang members' profiles. Word comparison shows that gang members prefer to use curse words ({\em nigga, fuck, shit}) in their profile descriptions while non-gang members use words related to their feelings or interests ({\em love, life, live, music, book}). The terms {\em rip} and {\em free} which appear in approximately $12\%$ of all gang member Twitter profiles, suggest that gang members use their profile descriptions as a space to grieve for their fallen or incarcerated gang members. The term {\em gang} in gang members' profile descriptions suggest that gang members like to self-identify themselves on Twitter. Such lexical features may therefore be of great importance for automatically identifying gang member profiles. We take counts of unigrams from gang and non-gang members' Twitter profile descriptions as classification features.

\begin{figure}
\centering
\includegraphics[width=0.95\linewidth]{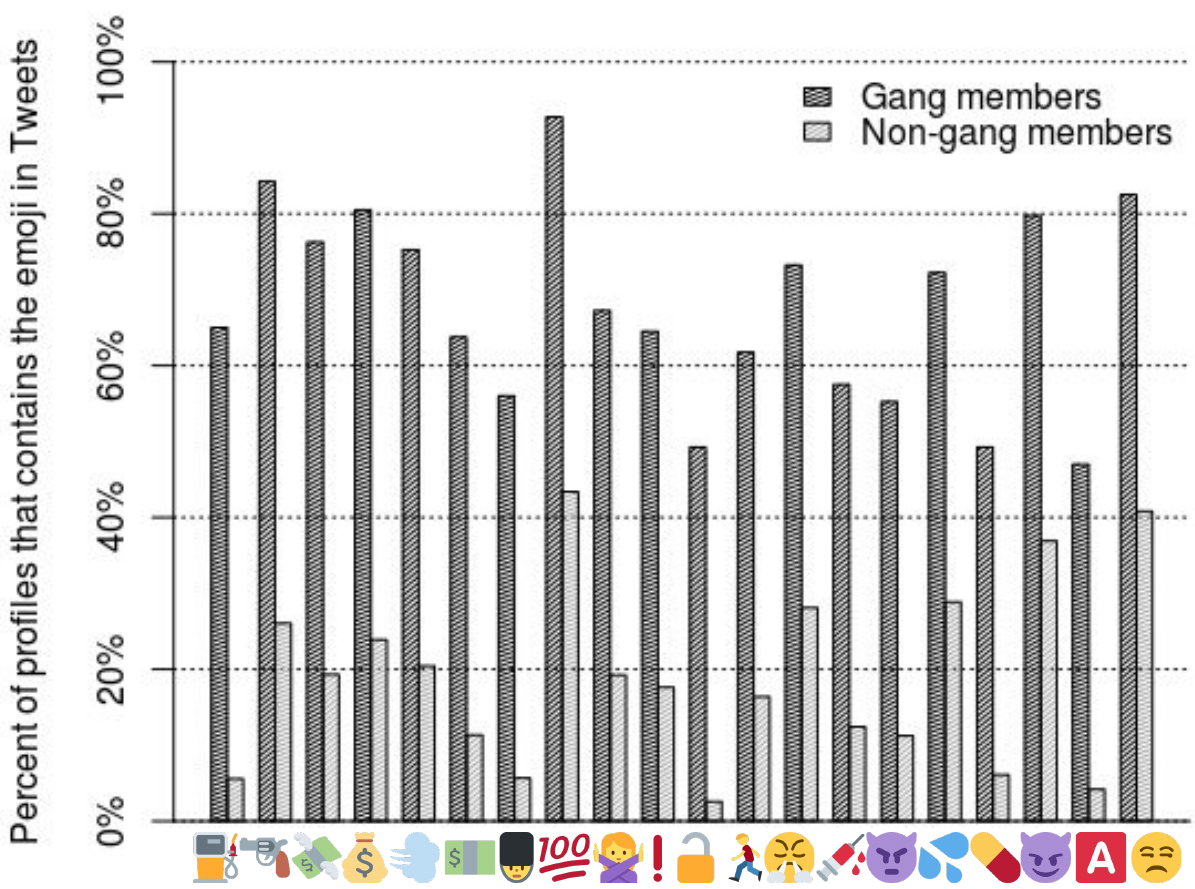} 
\caption{Emoji usage distribution: gang vs non-gang.}
\label{emojichart}
\vspace{-20px}
\end{figure}

\subsubsection{Music interests}  
It has been recognized that music is a key cultural component in an urban lifestyle and that gang members often want to emulate the scenarios and activities the music conveys~\cite{patton13}. Our analysis confirms that the influence of gangster rap is expressed in gang members' Twitter posts. We found that 51.25\% of the gang members collected have a tweet that links to a YouTube video. Following these links, a simple keyword search for the terms \texttt{gangsta} and \texttt{hip-hop} in the YouTube video description found that 76.58\% of the shared links are related to hip-hop music, gangster rap, and the culture that surrounds this music genre. Moreover, this high proportion is not driven by a small number of profiles that prolifically share YouTube links; eight YouTube links are shared on average by a gang member.

Recognizing the frequency with which gang members post YouTube links on gangster rap and hip-hop, we consider the YouTube videos posted in a user's tweets as features for the classifier. In particular, for each YouTube video tweeted, we used the YouTube API\footnote{\url{https://developers.google.com/youtube/}} to retrieve the video's description and its comments. Further analysis of YouTube data showed a difference between terms in gang members' YouTube data and non-gang members' YouTube data. For example, the top 5 terms (after stemming and stop word removal) used in YouTube videos shared by  gang members are {\em shit, like, nigga, fuck, lil} while {\em like, love, peopl, song, get} are the top 5 terms in non-gang member video data. To represent a user profile based on their music interests, we generated a bag of words from the video descriptions and comments from all shared videos.

\subsubsection{Emoji}
\begin{figure}
\centering
\includegraphics[width=1\linewidth]{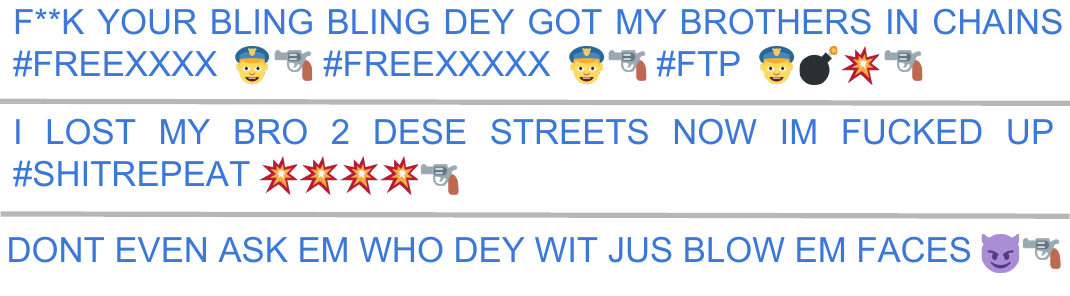} 
\caption{Examples for gang members' tweets with emojis.}
\label{fig:image3}
\vspace{-20px}
\end{figure}

Motivated by recent work involving the use of emojis by gang members~\cite{patton2016gang}, we also studied if and how gang and non-gang members use emoji symbols in their tweets. Our analysis found that gang members have a penchant for using just a small set of emoji symbols that convey their anger and violent behavior through their tweets. Figure~\ref{emojichart} illustrates the emoji distribution for the top 20 most frequent emojis used by gang member profiles in our dataset. The fuel pump emoji was the most frequently used emoji by the gang members, which is often used in the context of selling or consuming marijuana. The pistol emoji is the second most frequent in our dataset, which is often used with the guardsman emoji or the police cop emoji in an `emoji chain'. Figure~\ref{fig:image3} presents some prototypical `chaining' of emojis used by gang members. The chains may reflect their anger at law enforcement officers, as a cop emoji is often followed up with the emoji of a weapon, bomb, or explosion. We found that 32.25\% of gang members in our dataset have chained together the police and the pistol emoji, compared to just 1.14\% of non-gang members. Moreover, only 1.71\% of non-gang members have used the hundred points emoji and pistol emoji together in tweets while 53\% of gang members have used them. A variety of the angry face emoji such as devil face emoji and imp emoji were also common in gang member tweets. The frequency of each emoji symbol used across the set of user's tweets are thus considered as features for our classifier.

\subsubsection{Profile image}
\begin{figure}
\centering
\includegraphics[width=1\linewidth]{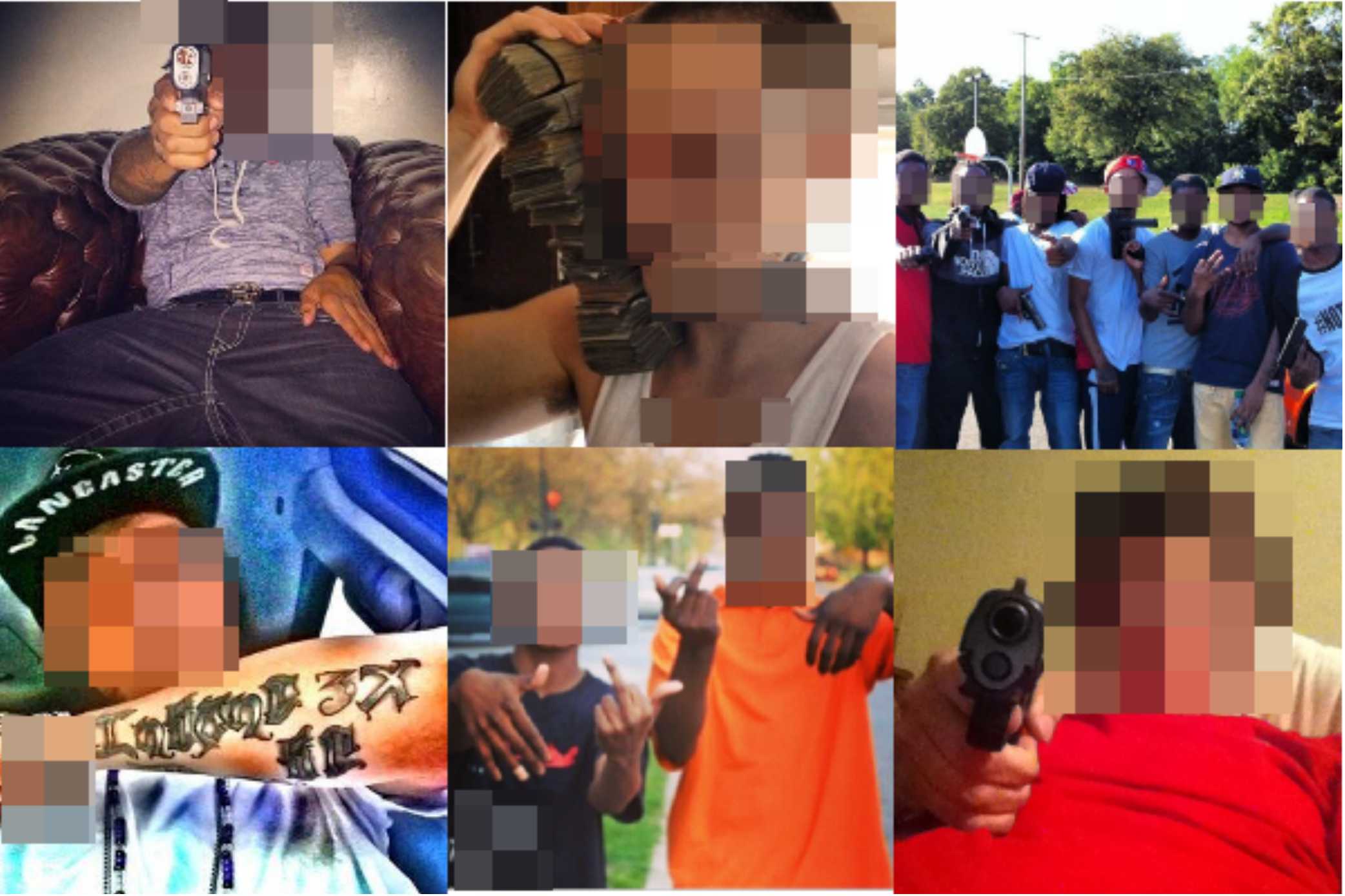} 
\caption{Sample gang member profile images.}
\label{gangprofpics}
\vspace{-20px}
\end{figure}

In our profile verification process, we observed that most gang member profiles portray a context representative of gang culture. Some examples of these profile pictures are shown in Figure~\ref{gangprofpics}, where the user holds or points weapons, is seen in a group fashion which displays a gangster culture, or is showing off graffiti, hand signs, tattoos and bulk cash. Descriptions of these images may thus empower our classifier. Thus, we translated profile images into features with the Clarifai web service\footnote{\url{http://www.clarifai.com/}}. Clarifai offers a free API to query a deep learning system that tags images with a set of scored keywords that reflect what is seen in the image. We tagged the profile image and cover image for each profile using 20 tags identified by Clarifai. Figure~\ref{fig:imagetags} offers the 20 most often used tags applied to gang and non-gang member profiles. Since we take all the tags returned for an image, we see common words such as {\em people and adult} coming up in the top 20 tag set. However, gang member profile images were assigned unique tags such as {\em trigger, bullet, worship} while non-gang images were uniquely tagged with {\em beach, seashore, dawn, wildlife, sand, pet}. The set of tags returned by Clarifai were thus considered as features for the classifier.

\subsection{Learning algorithms}\label{sec:la}
The unigrams of tweets, profile text, and linked YouTube video descriptions and comments, along with the distribution of emoji symbols and the profile image tags were used to train four different classification models: a Naive Bayes net, a Logistic Regression, a Random Forest, and a Support Vector Machine (SVM). These four models were chosen because they are known to perform well over text features, which is the dominant type of feature considered. The performance of the models are empirically compared to determine the most suitable classification technique for this problem. Data for the models are represented as a vector of term frequencies where the terms were collected from one or more feature sets described above.

\section{Evaluation} \label{sec:eval}
We next evaluate the performance of classifiers that use the above features to discover gang member profiles on Twitter. For this purpose, we use the training set discussed in Section~\ref{sec:dc} with 400 gang member profiles (the `positive'/`gang' class) and 2,865 non-gang member profiles (the `negative'/`non-gang' class). We trained and evaluated the performance of the classifiers mentioned in Section~\ref{sec:la} under a 10-fold cross validation scheme. For each of the four learning algorithms, we consider variations involving only tweet text, emoji, profile, image, or music interest (YouTube comments and video description) features, and a final variant that considers all types of features together. The classifiers that use a single feature type were intended to help us study the quality of their predictive power by itself. When building these single-feature classifiers, we filtered the training dataset based on the availability of the single feature type in the training data. For example, we only used the twitter profiles that had at least a single emoji in their tweets to train classifiers that consider emoji features. We found 3,085 such profiles out of the 3,265 profiles in the training set. When all feature types were considered, we developed two different models: 

\begin{enumerate}
\item {\em Model(1)}: This model is trained with all profiles in the training set. 
\item {\em Model(2)}: This model is trained with profiles that contain {\em every} feature type.
\end{enumerate}
Because a Twitter profile may not have every feature type, {\em Model(1)} represents a practical scenario where not every Twitter profile contains every type of feature. In this model, the non-occurrence of a feature is represented by `zeroing out' the feature value during model training. {\em Model(2)} represents the ideal scenario where all profiles contain {\em every} feature type. For this model, we used 1,358 training instances (42\% of all training instances), out of which 172 were gang members (43\% of all gang members) and 1,186 were non-gang members (41\% of all non-gang members). We used version 0.17.1 of scikit-learn\footnote{\url{http://scikit-learn.org/stable/index.html}} machine learning library to implement the classifiers.

\begin{figure}
\centering
\includegraphics[width=1\linewidth]{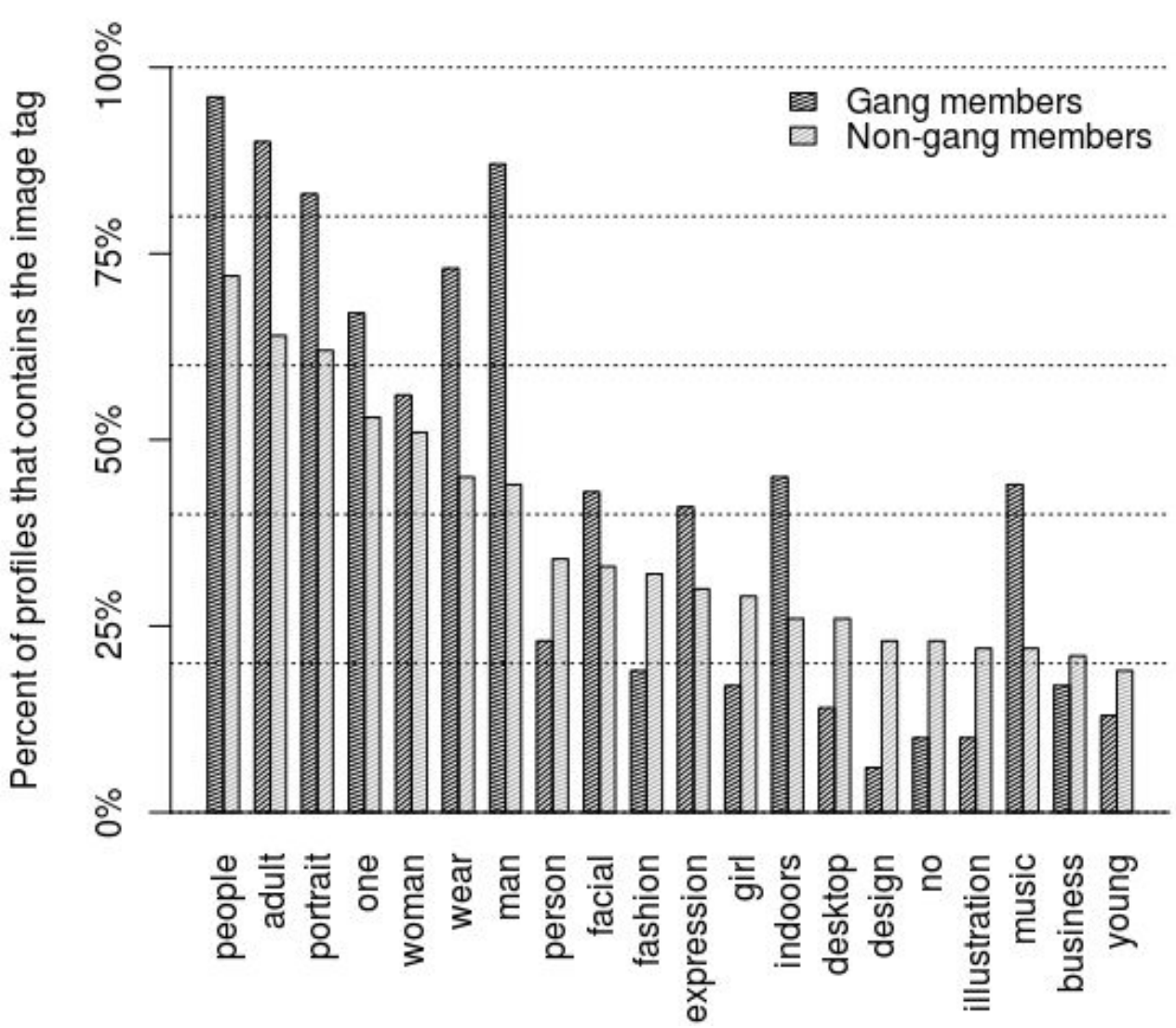} 
\caption{Image tags distribution: gang vs non-gang.}
\label{fig:imagetags}
\vspace{-20px}
\end{figure}

\begin{table*}[]
\centering
\begin{tabular}{|l|c|l|c|c|c|c|c|c|}
\hline
\multicolumn{1}{|c|}{\multirow{3}{*}{\textbf{Features}}} & \multirow{3}{*}{\begin{tabular}[c]{@{}c@{}}\textbf{Total Number of Profiles}\\ \textbf{\{\#Gang : \#Non-Gang\}}\end{tabular}} & \multicolumn{1}{c|}{\multirow{3}{*}{\textbf{Classifier}}} & \multicolumn{6}{c|}{\textbf{Results}}                                  \\ \cline{4-9} 
\multicolumn{1}{|c|}{}                          &                                                                                              & \multicolumn{1}{c|}{}                            & \multicolumn{3}{c|}{\textbf{Gang}} & \multicolumn{3}{c|}{\textbf{Non-Gang}} \\ \cline{4-9} 
\multicolumn{1}{|c|}{}                          &                                                                                              & \multicolumn{1}{c|}{}                              & \textbf{Precision} & \textbf{Recall} &  \textbf{$F1$-score}   & \textbf{Precision} & \textbf{Recall} &  \textbf{$F1$-score} \\ \hline
\multirow{3}{*}{Tweets (T)}                     & \multirow{3}{*}{3,265 \{400 : 2,865\}}                                                              & Naive Bayes                                         & 0.4354     & \textbf{0.9558}  & 0.5970      & \textbf{0.9929}     & 0.8278 & 0.9028  \\ 
                                         \cline{3-9} 
                                                &                                                                                              & Logistic Regression                                        & 0.6760     & 0.6623 & 0.6666     & 0.9529     & 0.9544 & 0.9536 \\
                                              \cline{3-9} 
                                                &                                                                                              & Random Forest        
                                                   & 0.8433     & 0.6401  & 0.7229     & 0.9517     & 0.9832 & 0.9671 \\
                                                
                                                \cline{3-9} 
                                                &                                                                                              & SVM        
                                                   & 0.6301    & 0.6545  & 0.6388      & 0.9514     & 0.9442 & 0.9477  \\
                                    
                                                \hline
\multirow{3}{*}{Emojis (E)}                     & \multirow{3}{*}{3,085 \{396 : 2,689\}}                                                              & Naive Bayes                                         & 0.4934     & 0.7989  & 0.6067     & 0.9676     & 0.8785  & 0.9207 \\ 
 \cline{3-9} 
                                                &                                                                                              & Logistic Regression                                        & 0.6867     & 0.3995 & 0.4969      & 0.9164     & 0.9733 & 0.9438 \\
 \cline{3-9} 
                                                &                                                                                              & Random Forest                                       & 0.7279     & 0.5079  & 0.5931     & 0.9292     & 0.9721  & 0.9500 \\ 
                                                
                                                \cline{3-9} 
                                                &                                                                                              & SVM        
                                                    & 0.4527     & 0.5642 & 0.4955     & 0.9329     & 0.8953 & 0.9133 \\

                                                \hline
\multirow{3}{*}{Profile data (P)}               & \multirow{3}{*}{2,996 \{378 : 2,618\}}                                                              & Naive Bayes                                           & 0.6000     & 0.243 & 0.464      & 0.8765     & \textbf{1.0000}  & 0.9341 \\ 
\cline{3-9}
                                                &                                                                                              & Logistic Regression                                      & 0.8015     & 0.2160  & 0.3362      & 0.8974     & 0.9924  & 0.9424 \\ 
 \cline{3-9} 
                                                &                                                                                              & Random Forest                                       & 0.5719     & 0.1441  & 0.2239     & 0.8886     & 0.9859 & 0.9346 \\ 
                                                
                                                \cline{3-9} 
                                                &                                                                                              & SVM        
                                                   & 0.7501     & 0.2225  & 0.3394     & 0.8978     & 0.9897  & 0.9414  \\

                                                \hline
\multirow{3}{*}{Image tags (I)}                 & \multirow{3}{*}{2,910 \{357 : 2,553\}}                                                              & Naive Bayes                                          & 0.2692     & 0.6973 & 0.3851    & 0.9458     & 0.7357 & 0.8271  \\ 
 \cline{3-9} 
                                                &                                                                                              & Logistic Regression                                        & 0.4832     & 0.1853 & 0.2624    & 0.8950     & 0.9722 & 0.9318  \\ 
 \cline{3-9} 
                                                &                                                                                              & Random Forest                                        & 0.4131     & 0.1512 & 0.2147    & 0.8911     & 0.9731 & 0.9300   \\ 
                                                
                                                \cline{3-9} 
                                                &                                                                                              & SVM        
                                                   & 0.3889     & 0.1454 & 0.205       & 0.8898     & 0.9679 & 0.9270  \\

                                                \hline
\multirow{3}{*}{Music interest (Y)}             & \multirow{3}{*}{1,630 \{196 : 1,434\}}                                                              & Naive Bayes                                          & 0.5865     & 0.7424 & 0.6505     & 0.9632     & 0.9297 & 0.9460 \\ 
 \cline{3-9} 
                                                &                                                                                              & Logistic Regression                                        & 0.7101     & 0.5447 & 0.6110      & 0.9395     & 0.9679 & 0.9534 \\ 
 \cline{3-9} 
                                                &                                                                                              & Random Forest                                        & 0.8403     & 0.3953 & 0.5277      & 0.9232     & 0.9895 & 0.9550 \\

                                                \cline{3-9} 
                                                &                                                                                              & SVM        
                                                    & 0.6232     & 0.6067 & 0.6072     & 0.9463     & 0.9476  & 0.9467\\

                                                \hline
\multirow{3}{*}{{\em Model(1)} \{T + E + P + I + Y\}}              & \multirow{3}{*}{3,265 \{400 : 2,865\}}                                                              & Naive Bayes                                          & 0.3718     & 0.9387 & 0.5312     & 0.9889     & 0.7791 & 0.8715  \\ 
\cline{3-9} 
                                                &                                                                                              & Logistic Regression                                        & 0.7250     & 0.6880 & 0.7038     & 0.9564     & 0.9637 & 0.9599 \\ 

 \cline{3-9} 
                                                &                                                                                              & Random Forest                                        & 0.8792     & 0.6374 & 0.7364     & 0.9507     & 0.9881 & 0.9690  \\
                                                
                                                \cline{3-9} 
                                                &                                                                                              & SVM        
                                                    & 0.6442     & 0.6791 & 0.6583     & 0.9546     & 0.9469 & 0.9506 \\

                                                \hline
\multirow{4}{*}{{\em Model(2)} \{T + E + P + I + Y\}}              & \multirow{4}{*}{1,358 \{172 : 1,186\}}                                                              & Naive Bayes                                          & 0.4405     & 0.9386 & 0.5926      & 0.9889      & 0.8254 & 0.8991 \\ 
 \cline{3-9} 
                                                &                                                                                              & Logistic Regression                                        & 0.7588     & 0.7396 & 0.7433  	& 0.9639 	& 0.9662 & 0.9649       \\ 
 
 \cline{3-9} 
                                                &                                                                                              & Random Forest                                   	& \textbf{0.8961}	& 0.6994 & \textbf{0.7755}		& 0.9575	& 0.9873 & \textbf{0.9720}  
                                                  \\
                                                  
                                                \cline{3-9} 
                                                &                                                                                              & SVM        
                                                    & 0.7185     & 0.7394 & 0.7213      & 0.9638    & 0.9586 & 0.9610  \\

                                                  \hline                                                
\end{tabular}
\caption{Classification results based on 10-fold cross validation.}
\label{results:usertweetsuni}
\vspace{-20px}
\end{table*}

For each 10-fold cross validation experiment, we report three evaluation metrics for the `gang' and `non-gang' classes, namely, the Precision = \(tp / (tp + fp)\), Recall = \(tp / (tp + fn)\), and $F1$-score = \(2 * (Precision * Recall) / (Precision + Recall) \),  where \(tp\) is the number of true positives, \(fp\) is the number of false positives, \(tn\) is the number of true negatives, and \(fn\) is the number of false negatives. We report these metrics for the positive `gang' and negative `non-gang' classes separately because of class imbalance in our dataset. 

\subsection{Experimental results}
Table~\ref{results:usertweetsuni} presents the average precision, recall, and $F1$-score over the 10 folds for the single-feature and combined feature classifiers. The table includes, in braces (`\{ \}'), the number of gang and non-gang profiles that contain a particular feature type, and hence the number of profiles used for the 10-fold cross validation. It is reasonable to expect that {\em any} Twitter profile is not that of a gang member, predicting a Twitter user as a non-gang member is much easier than predicting a Twitter user as a gang member. Moreover false positive classifications of the `gang' class may be detrimental 
to law enforcement investigations, which may go awry as they surveil an innocent person based on the classifier's suggestion. We thus believe that a small false positive rate of the `gang' class to be an especially important evaluation metric. We say that a classifier is `ideal' if it demonstrates high precision, recall, and $F1$-score for the `gang' class while performing well on the `non-gang' class as well.

The best performing classifier that considers single features is a Random Forest model over tweet features (T), with a reasonable $F1$-score of 0.7229 for the `gang' class. It also features the highest $F1$-score for the `non-gang' class (0.9671). Its strong performance is intuitive given the striking differences in language as shown in Figure~\ref{fig:image2} and discussed in Section~\ref{sec:data_analysis_tweet_text}. We also noted that music features offer promising results, with an $F1$-score of 0.6505 with a Naive Bayes classifier, as well as emoji features with an $F1$-score of 0.6067 also achieved by a Naive Bayes classifier. However, the use of profile data and image tags by themselves yield relatively poor $F1$-scores no matter which classifier considered. There may be two reasons for this despite the differences we observed in Section~\ref{sec:data_analysis}. First, these two feature types did not generate a large number of specific features for learning. For example,  descriptions are limited to just 160 characters per profile, leading to a limited number of unigrams (in our dataset, 10 on average) that can be used to train the classifiers. Second, the profile images were tagged by a third party Web service which is not specifically designed to identify gang hand signs, drugs and guns, which are often shared by gang members. This led to a small set of image tags in their profiles that were fairly generic, i.e., the image tags in Figure~\ref{fig:imagetags} such as `people', `man', and `adult'.

Combining these diverse sets of features into a single classifier yields even better results. Our results for {\em Model(1)} show that the Random Forest achieves the highest $F1$-scores for both `gang' (0.7364) and `non-gang' (0.9690) classes {\em and} yields the best precision of 0.8792, which corresponds to a low false positive rate when labeling a profile as a gang member. Despite the fact that it has lower positive recall compared to the second best performing classifier (a Random Forest trained over only tweet text features (T)), for this problem setting, we should be willing to increase the chance that a gang member will go unclassified if it means reducing the chance of applying a `gang' label to a non-gang member. When we tested {\em Model(2)}, a Random Forrest classifier achieved an $F1$-score of 0.7755 (improvement of 7.28\% with respect to the best performing single feature type classifier (T)) for `gang' class with a precision of 0.8961 (improvement of 6.26\% with respect to (T)) and a recall of 0.6994 (improvement of 9.26\% with respect to (T)). {\em Model(2)} thus outperforms {\em Model(1)}, and we expect its performance to improve with the availability of more training data with all feature types.

\vspace{-2px}
\subsection{Evaluation Over Unseen Profiles}
We also tested the trained classifiers using a set of Twitter profiles from a separate data collection process that may emulate the classifier's operation in a real-time setting. For this experiment, we captured real-time tweets from Los Angeles, CA\footnote{\url{http://isithackday.com/geoplanet-explorer/index.php?woeid=2442047}} and from ten South Side, Chicago neighborhoods that are known for gang-related activities~\cite{7165945} using the Twitter streaming API. We consider these areas with known gang presence on social media to ensure that some positive profiles would appear in our test set. We ultimately collected 24,162 Twitter profiles: 15,662 from Los Angeles, and 8,500 from Chicago. We populated data for each profile by using the 3,200 most recent tweets (the maximum that can be collected from Twitter's API) for each profile. Since the 24,162 profiles are far too many to label manually, we qualitatively study those profiles the classifier placed into the `gang' class.

\begin{figure}
\centering
\includegraphics[width=1\linewidth]{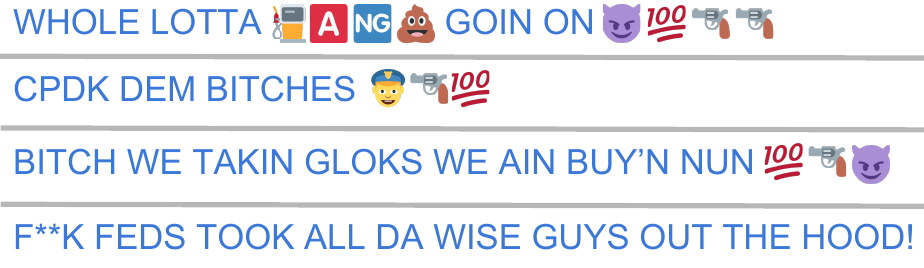} 
\caption{Sample tweets from identified gang members.}
\label{testset_examples}
\vspace{-20px}
\end{figure}

We used the training dataset to train our best performing random forest classifier (which use all feature types) and tested it on the test dataset. We then analyzed the Twitter profiles that our classifier labeled as belonging to the `gang' class. Each of those profiles had several features which overlap with gang members such as displaying hand signs and weapons in their profile images or in videos posted by them, gang names or gang-related hashtags in their profile descriptions, frequent use of curse words, and the use of terms such as {\em ``my homie"} to refer to self-identified gang members. Representative tweets extracted from those profiles are depicted in Figure~\ref{testset_examples}. The most frequent words found in tweets from those profiles were {\em shit, nigga, got, bitch, go, fuck etc.} and their user profiles had terms such as {\em free, artist, shit, fuck, freedagang, and ripthefallen}. They had frequently used emojis such as {\em face with tears of joy, hundred points symbol, fire, skull, money bag, and pistol}. For some profiles, it was less
obvious that the classifier correctly identified a gang member. Such profiles used the same emojis and curse words commonly found in gang members profiles, but their profile picture and tweet content was not indicative of a 
gang affiliation. In conclusion, we find that in a real-time-like setting, the classifier to be able to extract profiles 
with features that strongly suggest gang affiliation. Of course, these 
profiles demand further investigation and extensive evidence from other sources in order to draw a concrete conclusion, 
especially in the context of a law enforcement investigation. We refrain from reporting any profile names or specific details about the profiles labeled as a `gang' member to comply with the applicable IRB governing this human subject research.

\vspace{-5px}
\section{Conclusion and Future Work} \label{sec:con} 
This paper presented an approach to address the problem of automatically identifying gang member profiles on Twitter. Despite the challenges in developing such automated systems, mainly due to difficulties in finding online gang member profiles for developing training datasets, we proposed an approach that uses features extracted from textual descriptions, emojis, images and videos shared on Twitter (textual features extracted from images, and videos). Exploratory analysis of these types of features revealed interesting, and sometimes striking differences in the ways gang and non-gang members use Twitter. Classifiers trained over features that highlight these differences, were evaluated under 10-fold cross validation. Our best classifier achieved a promising $F1$-score of 0.7755 over the `gang' profiles when all types of features were considered.

Future work will strengthen our training dataset by including more gang member Twitter profiles by searching for more location-independent keywords. We also plan to develop our own image classification system specifically designed to classify images found on gang member profiles. We would also like to experiment with building dictionaries that contain gang names to understand whether {\em{``having a gang name in the profile description''}} as a feature can improve our results. Finally, we would also like to study how can we further improve our classifier models using word embeddings~\cite{gangwordembeddings} and social networks of known gang members.

\vspace{-5px}
\section{Acknowledgement} 

We are thankful to Uday Kiran Yeda for helping us with data collection. We acknowledge partial support from the National Science Foundation (NSF) award: CNS-1513721: ``Context-Aware Harassment Detection on Social Media'', National Institutes of Health (NIH) award: MH105384-01A1: ``Modeling Social Behavior for Healthcare Utilization in Depression'' and Grant No. 2014-PS-PSN-00006 awarded by the Bureau of Justice Assistance. The Bureau of Justice Assistance is a component of the U.S. Department of Justice's Office of Justice Programs, which also includes the Bureau of Justice Statistics, the National Institute of Justice, the Office of Juvenile Justice and Delinquency Prevention, the Office for Victims of Crime, and the SMART Office. Points of view or opinions in this document are those of the authors and do not necessarily represent the official position or policies of the U.S. Department of Justice, NSF or NIH.


\vspace{-5px}
\bibliographystyle{IEEEtran}
\input{gang16.bbl}


\end{document}

%% file: gang16.bbl

%% file: gang16.bbl
\begin{thebibliography}{10}
\providecommand{\url}[1]{#1}
\csname url@samestyle\endcsname
\providecommand{\newblock}{\relax}
\providecommand{\bibinfo}[2]{#2}
\providecommand{\BIBentrySTDinterwordspacing}{\spaceskip=0pt\relax}
\providecommand{\BIBentryALTinterwordstretchfactor}{4}
\providecommand{\BIBentryALTinterwordspacing}{\spaceskip=\fontdimen2\font plus
\BIBentryALTinterwordstretchfactor\fontdimen3\font minus
  \fontdimen4\font\relax}
\providecommand{\BIBforeignlanguage}[2]{{%
\expandafter\ifx\csname l@#1\endcsname\relax
\typeout{** WARNING: IEEEtran.bst: No hyphenation pattern has been}%
\typeout{** loaded for the language `#1'. Using the pattern for}%
\typeout{** the default language instead.}%
\else
\language=\csname l@#1\endcsname
\fi
#2}}
\providecommand{\BIBdecl}{\relax}
\BIBdecl

\bibitem{2011NationalGangThreat}
\emph{2011 National Gang Threat Assessment Issued Emerging Trends}, 2011.

\bibitem{2013NationalGangReport}
\emph{National Gang Report, National Gang Intelligence Center}, 2013.

\bibitem{miller1992crime}
W.~B. Miller, \emph{Crime by youth gangs and groups in the United
  States}.\hskip 1em plus 0.5em minus 0.4em\relax US Department of Justice,
  Office of Justice Programs, Office of Juvenile Justice and Delinquency
  Prevention Washington, DC, 1992.

\bibitem{centers2012gang}
``Gang homicides-five us cities, 2003-2008.'' \emph{Morbidity and mortality
  weekly report}, vol.~61, no.~3, pp. 46--51, 2012.

\bibitem{2007assesment}
\emph{Survey of Gang Members’ Online Habits and Participation (2007) Survey
  results reported at the i-SAFE Annual Internet Safety Education Review
  Meeting Carlsbad, California}.\hskip 1em plus 0.5em minus 0.4em\relax
  National Assessment Center.

\bibitem{King2007S66}
J.~E. King, C.~E. Walpole, and K.~Lamon, ``Surf and turf wars online—growing
  implications of internet gang violence,'' \emph{Journal of Adolescent
  Health}, vol.~41, no. 6, Supplement, pp. S66 -- S68, 2007.

\bibitem{doi:10.1080/07418825.2013.778326}
D.~C. Pyrooz, S.~H. Decker, and R.~K.~M. Jr., ``Criminal and routine activities
  in online settings: Gangs, offenders, and the internet,'' \emph{Justice
  Quarterly}, vol.~32, no.~3, pp. 471--499, 2015.

\bibitem{patton13}
D.~U. Patton, R.~D. Eschmann, and D.~A. Butler, ``Internet banging: New trends
  in social media, gang violence, masculinity and hip hop,'' \emph{Computers in
  Human Behavior}, vol.~29, no.~5, pp. A54 -- A59, 2013.

\bibitem{howell2010gang}
J.~C. Howell, ``Gang prevention: An overview of research and programs. juvenile
  justice bulletin.'' \emph{Office of Juvenile Justice and Delinquency
  Prevention}, 2010.

\bibitem{ito10}
M.~Ito, S.~Baumer, M.~Bittanti, R.~Cody, B.~Herr-Stephenson, H.~A. Horst, P.~G.
  Lange, D.~Mahendran, K.~Z. Mart{\'\i}nez, C.~Pascoe \emph{et~al.}, ``Hanging
  out, messing around, and geeking out,'' \emph{Digital media}, 2010.

\bibitem{7165945}
S.~Wijeratne, D.~Doran, A.~Sheth, and J.~L. Dustin, ``Analyzing the social
  media footprint of street gangs,'' in \emph{IEEE International Conference on
  Intelligence and Security Informatics (ISI), 2015}, May 2015, pp. 91--96.

\bibitem{police13}
P.~E.~R. Forum, ``Social media and tactical considerations for law
  enforcement,'' United States Office of Community Oriented Policing Services
  and United States Department of Justice, Tech. Rep., 2013.

\bibitem{brunty14}
J.~Brunty, L.~Miller, and K.~Helenek, \emph{Social media investigation for law
  enforcement}.\hskip 1em plus 0.5em minus 0.4em\relax Routledge, 2014.

\bibitem{pennacchiotti2011machine}
M.~Pennacchiotti and A.-M. Popescu, ``A machine learning approach to twitter
  user classification,'' 2011.

\bibitem{tinati2012identifying}
R.~Tinati, L.~Carr, W.~Hall, and J.~Bentwood, ``Identifying communicator roles
  in twitter,'' in \emph{Proceedings of the 21st International Conference on
  World Wide Web}, ser. WWW '12 Companion.\hskip 1em plus 0.5em minus
  0.4em\relax New York, NY, USA: ACM, 2012, pp. 1161--1168.

\bibitem{liu2013s}
W.~Liu and D.~Ruths, ``What’s in a name? using first names as features for
  gender inference in twitter,'' 2013.

\bibitem{purohit2012user}
H.~Purohit, A.~Dow, O.~Alonso, L.~Duan, and K.~Haas, ``User taglines:
  Alternative presentations of expertise and interest in social media,'' in
  \emph{2012 International Conference on Social Informatics (Social
  Informatics), Washington, D.C., USA, December 14-16}, 2012, pp. 236--243.

\bibitem{thrasher1963gang}
F.~M. Thrasher, \emph{The gang: A study of 1,313 gangs in Chicago}.\hskip 1em
  plus 0.5em minus 0.4em\relax University of Chicago Press, 1963.

\bibitem{patton2014social}
D.~U. Patton, J.~S. Hong, M.~Ranney, S.~Patel, C.~Kelley, R.~Eschmann, and
  T.~Washington, ``Social media as a vector for youth violence: A review of the
  literature,'' \emph{Computers in Human Behavior}, 2014.

\bibitem{decker2011leaving}
S.~Decker and D.~Pyrooz, ``Leaving the gang: Logging off and moving on. council
  on foreign relations,'' 2011.

\bibitem{desmondupatton2015}
D.~U. Patton, ``Gang violence, crime, and substance use on twitter: A snapshot
  of gang communications in detroit,'' Society for Social Work and Research
  19th Annual Conference: The Social and Behavioral Importance of Increased
  Longevity, jan 2015.

\bibitem{wang2014cursing}
W.~Wang, L.~Chen, K.~Thirunarayan, and A.~P. Sheth, ``Cursing in english on
  twitter,'' in \emph{Proceedings of the 17th ACM Conference on Computer
  Supported Cooperative Work \& Social Computing}, ser. CSCW '14.\hskip 1em
  plus 0.5em minus 0.4em\relax New York, NY, USA: ACM, 2014, pp. 415--425.

\bibitem{patton2016gang}
D.~U. Patton, J.~Lane, P.~Leonard, J.~Macbeth, and J.~R. Smith-Lee, ``Gang
  violence on the digital street: Case study of a south side chicago gang
  member’s twitter communication,'' \emph{New Media \& Society}, 2016.

\bibitem{gangwordembeddings}
S.~Wijeratne, L.~Balasuriya, D.~Doran, and A.~Sheth, ``Word embeddings to
  enhance twitter gang member profile identification,'' in \emph{IJCAI Workshop
  on Semantic Machine Learning (SML 2016)}.\hskip 1em plus 0.5em minus
  0.4em\relax New York City, NY: CEUR-WS, 07/2016 2016.

\end{thebibliography}
